\documentclass[conference]{IEEEtran}
\IEEEoverridecommandlockouts
\pdfoutput=1
\usepackage{cite}
\usepackage{amsmath,amssymb,amsfonts}
\usepackage{algorithmic}
\usepackage{graphicx}
\usepackage{textcomp}
\usepackage{xcolor}
\usepackage{fancyhdr}
\def\BibTeX{{\rm B\kern-.05em{\sc i\kern-.025em b}\kern-.08em
    T\kern-.1667em\lower.7ex\hbox{E}\kern-.125emX}}

\usepackage{dblfloatfix}  
\usepackage[braket]{qcircuit}
\usepackage{physics}

\graphicspath{{svg-inkscape/}}

\usepackage{pgfplots}
\pgfplotsset{compat=1.14}

\usepackage[final]{changes}  

\newcommand{\addcolor}{blue!70!green}
\setaddedmarkup{\textcolor{\addcolor}{\uwave{#1}}}
\setdeletedmarkup{\textcolor{red!70!black}{\sout{#1}}}

\setdeletedmarkup{}

\newcommand{\hide}[1]{}

\newcommand\figscale{1}
\newcommand\figfirstextrascale{1}
\newcommand\figcompactextrascale{1}
\newcommand\figcnotextrascale{1}
\newcommand\figscaleplot{1}
\newcommand\sensitivitytwowidth{0.322}

\fancypagestyle{firststyle}{%
  \fancyhf{}%
  
  \fancyfoot[C]{\scriptsize\vspace{-2em}%
    \copyright 2020 IEEE.  Personal use of this material is permitted.  Permission from IEEE must be obtained for all other uses, in any current or future media, including reprinting/republishing this material for advertising or promotional purposes, creating new collective works, for resale or redistribution to servers or lists, or reuse of any copyrighted component of this work in other works.%
  }%
}
\begin{document}

\title{Virtualized Logical Qubits: A 2.5D Architecture for Error-Corrected Quantum Computing
\thanks{This work is funded in part by EPiQC, an NSF Expedition in Computing, under grants CCF-1730449/1832377; in part by STAQ under grant NSF PHY-1818914; and in part by DOE grants DE-SC0020289 and DE-SC0020331.}
}

\author{
\IEEEauthorblockN{Casey Duckering, Jonathan M. Baker, David I. Schuster, and Frederic T. Chong\hspace{0.8em}}
\IEEEauthorblockA{
\textit{University of Chicago\hspace{0.8em}}\\
\{cduck, jmbaker, david.schuster, ftchong\}@uchicago.edu\hspace{0.8em}}
}

\maketitle
\thispagestyle{firststyle}

\begin{abstract}

Current, near-term quantum devices have shown great progress in the last several years culminating recently with a demonstration of quantum supremacy.  In the medium-term, however, quantum machines will need to transition to greater reliability through error correction, likely through promising techniques like surface codes which are well suited for near-term devices with limited qubit connectivity.  We discover quantum memory, particularly resonant cavities with transmon qubits arranged in a 2.5D architecture, can efficiently implement surface codes with substantial hardware savings and performance/fidelity gains.  Specifically, we {\em virtualize logical qubits} by storing them in layers of qubit memories connected to each transmon.

Surprisingly, distributing each logical qubit across many memories has a minimal impact on fault tolerance and results in substantially more efficient operations.  Our design permits fast transversal application of CNOT operations between logical qubits sharing the same physical address (same set of cavities) which are 6x faster than standard lattice surgery CNOTs.  We develop a novel embedding which saves approximately 10x in transmons with another 2x savings from an additional optimization for compactness.

Although qubit virtualization pays a 10x penalty in serialization, advantages in the transversal CNOT and in area efficiency result in fault-tolerance and performance comparable to conventional 2D transmon-only architectures.  Our simulations show our system can achieve fault tolerance comparable to conventional two-dimensional grids while saving substantial hardware.  Furthermore, our architecture can produce magic states at 1.22x the baseline rate given a fixed number of transmon qubits.  This is a critical benchmark for future fault-tolerant quantum computers as magic states are essential and machines will spend the majority of their resources continuously producing them.  This architecture substantially reduces the hardware requirements for fault-tolerant quantum computing and puts within reach a proof-of-concept experimental demonstration of around 10 logical qubits, requiring only 11 transmons and 9 attached cavities in total.

\end{abstract}


\section{Introduction}
\label{sec:introduction}

Quantum devices have improved significantly in the last several years both in terms of physical error rates and number of usable qubits. For example, IBM and others have made accessible via the cloud several devices with 5 to 53 qubits with moderate error rates \cite{ibm-device}. Concurrently, great progress has been made at the software level such as improved compilation procedures reducing required overhead for program execution. These efforts are directed at enabling NISQ (Noisy Intermediate-Scale Quantum) \cite{nisq} algorithms to demonstrate the power of quantum computing. Machines in this era are expected to run some important programs and have recently been used to by Google to demonstrate ``quantum supremacy'' \cite{quantum-supremacy}.

Despite this, these machines will be too small for error correction and unable to run large-scale programs due to unreliable qubits. The ultimate goal is to construct fault-tolerant machines capable of executing thousands of gates and in the long-term to execute large-scale  algorithms such as Shor's \cite{shor} and Grover's \cite{grover} with speedups over classical algorithms. There are a number of promising error correction schemes which have been proposed such as the color code \cite{color-code} or the surface code \cite{fowler-braid,fowler-lattice,gidney2019factor}. The surface code is a particularly appealing candidate because of its low overhead, high error threshold, and its reliance on few nearest-neighbor interactions in a 2D array of qubits, a common feature of superconducting transmon qubit hardware. In fact, Google's next milestone is to demonstrate error corrected qubits \cite{quantum-supremacy,martinis-caltech}.

Current architectures for both NISQ and fault-tolerant quantum computers make no distinction between the memory and processing of quantum information (represented in qubits). While currently viable, as larger devices are built, the engineering challenges of scaling up to hundreds of qubits becomes readily apparent. For transmon technology used by Google, IBM, and Rigetti, some of these issues include fabrication consistency and crosstalk during parallel operations. Every qubit needs dedicated control wires and signal generators which fill the refrigerator the device runs in.
To scale to the millions of qubits needed for useful fault-tolerant machines \cite{gidney2019factor}, we need to a memory-based architecture to decouple qubit-count from transmon-count.

In this work, we use a recently realized qubit memory technology which stores qubits in a superconducting cavity \cite{schuster-arch}. This technology, while new, is expected to become competitive with existing transmon devices. Stored in cavity, qubits have a significantly longer lifetime (coherence time) but must be loaded into a transmon for computation.  Although the basic concept of a compute qubit and associated memory has been demonstrated experimentally, the contribution of our work is to design and evaluate a system-level organization of these components within the context of a novel surface code embedding and fault-tolerant quantum operations.
We provide a proof of concept in the form of a practical use case motivating more complex experimental demonstrations of larger systems using this technology.

\begin{figure}[t]
    \centering
    \scalebox{\figfirstextrascale}{%
    \scalebox{\figscale}{%
    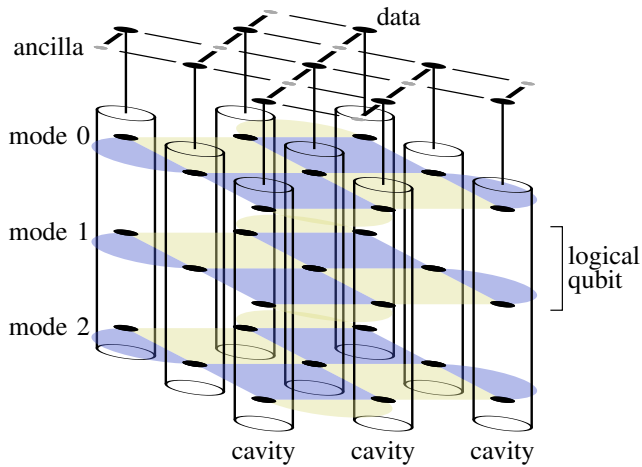%
    }}
    \caption{Our fault-tolerant architecture with random-access memory local to each transmon.  On top is the typical 2D grid of transmon qubits.  Attached below each data transmon is a resonant cavity storing error-prone data qubits (shown as black circles).
    This pattern is tiled in 2D to obtain a 2.5D array of logical qubits.
    Our key innovation here is storing the qubits that make up each logical qubit (shown as checkerboards) across many cavities to enable efficient computation.
    }
    \label{fig:stacked-surfaces}
\end{figure}

Our proposed 2.5D memory-based design is a typical 2D grid of transmons with memory added as shown in Figure~\ref{fig:stacked-surfaces}. This can be compared with the traditional 2D error correction implementation in Figure~\ref{fig:regular-surface-code}, where the checkerboards represent error-corrected logical qubits. The logical qubits in this system are stored at unique virtual addresses in memory cavities when not in use. They are loaded to a physical address in the transmons and made accessible for computation on request and are periodically loaded to correct errors, similar to DRAM refresh. This design allows for more efficient operations such as the transversal CNOT between logical qubits sharing the same physical address i.e.\ co-located in the same cavities. This is not possible on the surface code in 2D which requires methods such as braiding or lattice surgery for a CNOT operation.

We introduce two embeddings of the 2D surface code to this new architecture that spread logical qubits across many cavities.
Despite serialization due to memory access, we are able to store and error-correct stacks of these logical qubits. Furthermore, we show surface code operations via lattice surgery can be used unchanged in this new architecture while also enabling a more efficient CNOT operation. Similarly, we are able to use standard and architecture-specific magic-state distillation protocols \cite{game-of-codes} in order to ensure universal computation. Magic-state distillation is a critical component of error-corrected algorithms so any improvement will directly speed up algorithms including Shor's and Grover's.

We discuss several important features of any proposed error correction code, such as the threshold error rate (below which the code is able to correct more errors than its execution causes), the code distance, and the number of physical qubits to encode a logical qubit. In many codes, the number of physical qubits can be quite large. We develop an embedding from the standard representation to this new architecture which reduces the required number of physical transmon qubits by a factor of approximately $k$, the number of resonant modes per cavity. We also develop a Compact variant saving an additional 2x. This is significant because we can obtain a code distance $\sqrt{2k}$ times greater or use hardware with only $\frac{1}{2k}$ the required physical transmons for a given algorithm. In the near-to-intermediate term, when qubits are a highly constrained resource this will accelerate a path towards fault-tolerant computation. In fact, the smallest instance of Compact requires only 11 transmons and 9 cavities for $k$ logical qubits.

We evaluate variants of our architecture by comparing against the surface code on a larger 2D device. Specifically, we determine the error correction threshold rates via simulation for each and find they are all close to the baseline threshold. This shows the additional error sources do not significantly impact the performance. We explore the sensitivity of the threshold to many different sources of error, some of which are unique to the memory used in this architecture.  We end by evaluating magic-state distillation protocols which have a large impact on overall algorithm performance and find a 1.22x speedup normalized by the number of transmon qubits.

In summary, we make the following contributions:
\begin{itemize}
    \item We introduce a 2.5D architecture where qubit-local memory is used for random access to error-corrected, logical qubits stored across different memories.  This allows a simple virtual and physical address scheme.  Error correction is performed continuously by loading each from memory.
    \item We give two efficient adaptations of the surface code in this architecture, Natural and Compact. Unlike a naive embedding, both support fast transversal CNOTs in addition to lattice surgery operations with improved connectivity between logical qubits.
    \item We develop an error correction implementation optimized for Compact and designed to maximise parallelism and minimize the spread of errors.
    \item Via simulation, we determine the surface code adapted to our 2.5D architecture is still an effective error correction code while greatly reducing hardware requirements.
\end{itemize}

\section{Background}
\label{sec:background}
In this section we briefly introduce the basics of quantum computation. We review current superconducting qubit architectures and memory technology our proposed design takes advantage of.  We then discuss the noise present in these physical systems. Next, we introduce the basics of quantum error correction and give a detailed introduction to the surface code and lattice surgery. We conclude with a review of the basic procedure for decoding physical errors.

\subsection{Basics of Quantum Computing}
\label{sec:basics}
The fundamental unit of quantum computing is the qubit. Like the classical bit, it can exist in the $\ket{0}$ or $\ket{1}$ state, but it may also exist in a coherent superposition of the two states and $n$ qubits may exist in a superposition of all $2^n$ bit strings. For example, a single qubit state is $\ket{\psi} = \alpha\ket{0} + \beta\ket{1}$ where $\abs{\alpha}^2 + \abs{\beta}^2 = 1$ and $\alpha, \beta \in \mathbb{C}$. To manipulate these bits we apply quantum operations, often called gates. Single qubit gates like X (bit flip), Z (phase flip), H (Hadamard basis change), and T ($\frac{\pi}{4}$ phase) and two-qubit gates like CNOT (reversible XOR with output $b'=a\oplus b$) are unitary and reversible (invertible). We may measure a qubit to obtain either a 0 or a 1 outcome with probabilities $\abs{\alpha}^2$ and $\abs{\beta}^2$, respectively. Multi-qubit operations like CNOT can create entanglement between qubits. Using only CNOT and single qubit gates, universal computation is possible, meaning any reversible multi-qubit operation is possible.  The three-qubit Toffoli (reversible AND gate with output $c'=(a\wedge b)\oplus c$), a common primitive in error-corrected algorithms, can be implemented by performing a few CNOT, H, and T gates.  See \cite{nielsen-chuang} for a more comprehensive background.

\subsection{Superconducting Qubit Architectures}
\label{sec:architectures}
In contrast to other leading qubit technologies such as trapped ion devices with one or more fully-connected qubit chains, superconducting qubits are typically connected in nearest-neighbor topologies, often a 2D mesh on a regular square grid. For near-term computation, this limitation makes engineering these devices easier but results in high communication costs, increasing the chance of errors on NISQ devices and communication congestion for error corrected operations.  This is a leading technology in industry, used by Rigetti, IBM, and Google.

\subsection{Qubit Memory Technology}
\begin{figure}[t]
    \centering
    \scalebox{\figfirstextrascale}{%
    \scalebox{\figscale}{%
\begingroup%
  \makeatletter%
  \providecommand\color[2][]{%
    \errmessage{(Inkscape) Color is used for the text in Inkscape, but the package 'color.sty' is not loaded}%
    \renewcommand\color[2][]{}%
  }%
  \providecommand\transparent[1]{%
    \errmessage{(Inkscape) Transparency is used (non-zero) for the text in Inkscape, but the package 'transparent.sty' is not loaded}%
    \renewcommand\transparent[1]{}%
  }%
  \providecommand\rotatebox[2]{#2}%
  \newcommand*\fsize{\dimexpr\f@size pt\relax}%
  \newcommand*\lineheight[1]{\fontsize{\fsize}{#1\fsize}\selectfont}%
  \ifx\svgwidth\undefined%
    \setlength{\unitlength}{104.02734512bp}%
    \ifx\svgscale\undefined%
      \relax%
    \else%
      \setlength{\unitlength}{\unitlength * \real{\svgscale}}%
    \fi%
  \else%
    \setlength{\unitlength}{\svgwidth}%
  \fi%
  \global\let\svgwidth\undefined%
  \global\let\svgscale\undefined%
  \makeatother%
  \begin{picture}(1,1.23130448)%
    \lineheight{1}%
    \setlength\tabcolsep{0pt}%
    \put(0,0){\includegraphics[width=\unitlength,page=1]{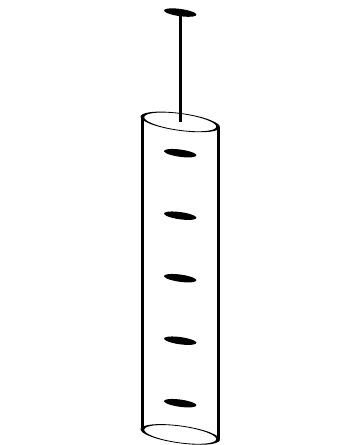}}%
    \put(0.62880196,0.7812369){\color[rgb]{0,0,0}\makebox(0,0)[lt]{\lineheight{1.25}\smash{\begin{tabular}[t]{l}mode 0\end{tabular}}}}%
    \put(0.73694661,0.41787092){\color[rgb]{0,0,0}\makebox(0,0)[lt]{\lineheight{1.25}\smash{\begin{tabular}[t]{l}\vdots\end{tabular}}}}%
    \put(0.62880196,0.08911121){\color[rgb]{0,0,0}\makebox(0,0)[lt]{\lineheight{1.25}\smash{\begin{tabular}[t]{l}mode k\end{tabular}}}}%
    \put(0.2827391,0.86775261){\color[rgb]{0,0,0}\makebox(0,0)[rt]{\lineheight{1.25}\smash{\begin{tabular}[t]{r}cavity\end{tabular}}}}%
    \put(0.41251267,1.17055761){\color[rgb]{0,0,0}\makebox(0,0)[rt]{\lineheight{1.25}\smash{\begin{tabular}[t]{r}transmon\end{tabular}}}}%
  \end{picture}%
\endgroup%
    }}
    \caption{A close-up representation of the qubit memory technology we use.  On top is a superconducting transmon qubit physically connected to a resonant superconducting cavity.  This cavity has many resonant modes each used to store a qubit.  These qubits can be loaded and stored (with random access) via the transmon.
    }
    \label{fig:single-cavity}
\end{figure}
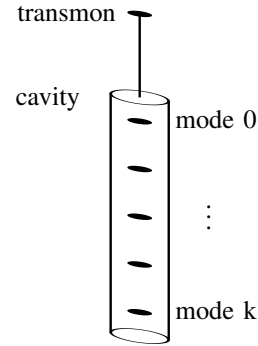

Recently, studies have demonstrated random access memory for quantum information \cite{schuster-arch,yale-arch}. Qubit states can be stored in the resonant modes of physical superconducting cavities attached to a transmon qubit as depicted in Figure~\ref{fig:single-cavity}. In these devices, transmon-transmon interactions are essentially the same as other superconducting transmon technology and transmon-cavity interactions are expected to perform similarly. Currently demonstrated error rates are promising, and there is nothing fundamental preventing this technology from becoming competitive with other transmon devices. We expect operation error rates to improve, cavity sizes and coherence times to increase and in general expect performance to improve as it has with other quantum technologies.

Local memory is not free. Stored qubits cannot be operated on directly. Instead, operations on this information are mediated through the transmon. Furthermore, to operate on qubits stored in memory, we first load the qubit from memory. Then we perform the desired operation on the transmons, and store the qubit back in its original location. A two-qubit operation such as a CNOT can also be performed directly between the transmon and a qubit in its connected cavity by manipulating higher states of the transmon.  We use this transmon-mode CNOT later.

\begin{figure}[t]
    \makebox[0.9\columnwidth][c]{\scalebox{\figscale}{
\begingroup%
  \makeatletter%
  \providecommand\color[2][]{%
    \errmessage{(Inkscape) Color is used for the text in Inkscape, but the package 'color.sty' is not loaded}%
    \renewcommand\color[2][]{}%
  }%
  \providecommand\transparent[1]{%
    \errmessage{(Inkscape) Transparency is used (non-zero) for the text in Inkscape, but the package 'transparent.sty' is not loaded}%
    \renewcommand\transparent[1]{}%
  }%
  \providecommand\rotatebox[2]{#2}%
  \newcommand*\fsize{\dimexpr\f@size pt\relax}%
  \newcommand*\lineheight[1]{\fontsize{\fsize}{#1\fsize}\selectfont}%
  \ifx\svgwidth\undefined%
    \setlength{\unitlength}{228.37499499bp}%
    \ifx\svgscale\undefined%
      \relax%
    \else%
      \setlength{\unitlength}{\unitlength * \real{\svgscale}}%
    \fi%
  \else%
    \setlength{\unitlength}{\svgwidth}%
  \fi%
  \global\let\svgwidth\undefined%
  \global\let\svgscale\undefined%
  \makeatother%
  \begin{picture}(1,0.79788527)%
    \lineheight{1}%
    \setlength\tabcolsep{0pt}%
    \put(0,0){\includegraphics[width=\unitlength,page=1]{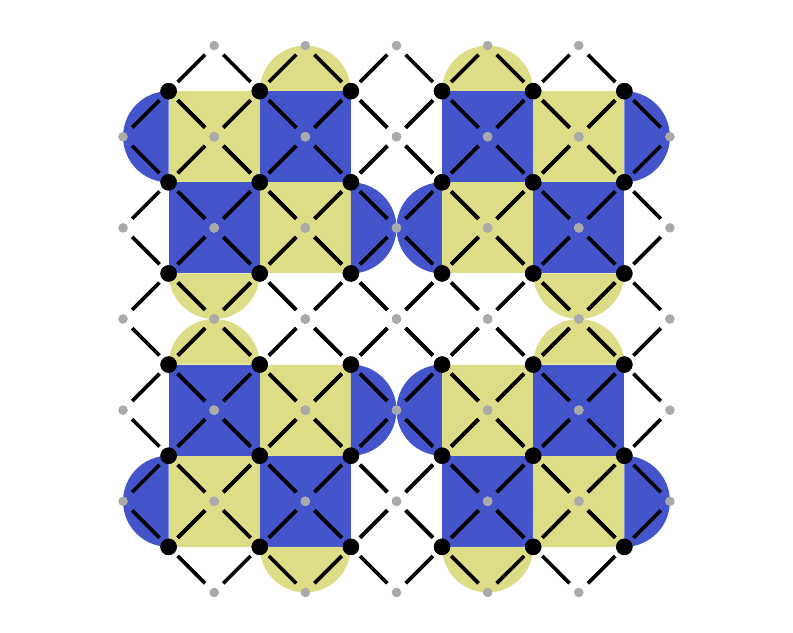}}%
    \put(0.79885058,0.69443699){\color[rgb]{0,0,0}\makebox(0,0)[lt]{\lineheight{1.25}\smash{\begin{tabular}[t]{l}data\end{tabular}}}}%
    \put(0.85632185,0.61857492){\color[rgb]{0,0,0}\makebox(0,0)[lt]{\lineheight{1.25}\smash{\begin{tabular}[t]{l}ancilla\end{tabular}}}}%
    \put(0.32758622,0.7634025){\color[rgb]{0,0,0}\makebox(0,0)[t]{\lineheight{1.25}\smash{\begin{tabular}[t]{c}logical qubit\end{tabular}}}}%
    \put(0.67241381,0.7634025){\color[rgb]{0,0,0}\makebox(0,0)[t]{\lineheight{1.25}\smash{\begin{tabular}[t]{c}logical qubit\end{tabular}}}}%
    \put(0.32758622,0.00478179){\color[rgb]{0,0,0}\makebox(0,0)[t]{\lineheight{1.25}\smash{\begin{tabular}[t]{c}logical qubit\end{tabular}}}}%
    \put(0.67241381,0.00478179){\color[rgb]{0,0,0}\makebox(0,0)[t]{\lineheight{1.25}\smash{\begin{tabular}[t]{c}logical qubit\end{tabular}}}}%
    \put(0.85632185,0.30823008){\color[rgb]{0,0,0}\makebox(0,0)[lt]{\lineheight{1.25}\smash{\begin{tabular}[t]{l}connectivity\end{tabular}}}}%
    \put(0,0){\includegraphics[width=\unitlength,page=2]{regular-surface-code.pdf}}%
  \end{picture}%
\endgroup%
}}
    \caption{A typical 2D superconducting qubit architecture.  The dots are transmon qubits where black are used as data and gray are used as ancilla for error correction.  The lines indicate physical connections between qubits that allow operations between them.  Four logical qubits, each consisting of 9 error-prone data qubits, are shown here in the rotated surface code with distance 3.  Z parity checks are shaded yellow (light) and X parity checks are shaded blue (dark) where checks on only 2 data are drawn as half circles.
    }
    \label{fig:regular-surface-code}
\end{figure}

In this architecture, qubits stored in the same cavity cannot be operated on in parallel. For example, consider two qubits stored in different modes of the same cavity (two virtual addresses corresponding to the same physical address). If we want to perform an H gate on each of them in parallel, this would not be possible. Instead, we serialize these operations. There are two primary benefits of this technology. First, we are able to quickly perform two-qubit interactions between any pair of qubits stored in the same cavity because we have star-graph connectivity between the transmon and its cavity modes. Second, qubits stored in the cavity are expected to have longer coherence times by about one order of magnitude i.e.\ there will be 10x fewer idle errors when qubits are stored in the cavity.

\subsection{Quantum Errors}
\label{sec:errors}

Quantum systems are inherently noisy, subject to a variety of coherent and non-coherent error. For example, when attempting to apply some gate $U$ to a qubit we may actually apply some other gate $U'$ which is close to the desired operation but may include an additional undesired operation. Fortunately, this type of coherent error is fairly easy to model. Since every single-qubit unitary can be expressed as a linear combination of the Pauli matrices\footnotemark $I, X, Y, Z$ we can express this coherent error as a combination of bit flip (X) and phase flip (Z) errors where $I$ is no error and $Y$ is simultaneous bit and phase errors ($Y=iXZ$). For a quantum error correcting code this will play a part in digitizing errors, meaning we will be able to simply detect and correct $X$ and $Z$ errors.
\footnotetext{The Pauli matrices $X=\begin{bmatrix}0&1\\1&0\end{bmatrix}$, $Y=\begin{bmatrix}0&-i\\i&0\end{bmatrix}$, $Z=\begin{bmatrix}1&0\\0&-1\end{bmatrix}$\\[2pt] along with $I$ form a complete basis over complex matrices so any single-qubit unitary $U=aI+bX+cY+dZ$.}

Errors such as decoherence errors can be attributed to interaction with the environment. These errors are inevitable because manipulating qubits requires they not be perfectly isolated. When modeling and simulating this type of error we require the use of full density matrix simulation. In this paper, we opt not to model coherence errors in this way because simulation of this class of errors is hard (density matrices have size exponential in the number of qubits), we instead also model storage errors as Pauli errors. This is a common simplification and a conservative overestimate for the error causing our error threshold estimation to be slightly more conservative. For example, when decoherence resets a qubit to $\ket{0}$, this causes an error to a qubit in the $\ket{1}$ state but not to a qubit already in the $\ket{0}$ state whereas a Pauli X error causes a bit flip which is an error on either state.

The above errors apply to all superconducting systems and we often assume consistent error rates across the device. We treat all two-qubit interactions equally so gates like a CNOT incur some fixed error cost, a fixed chance of some error $U_1 \otimes U_2$ is applied to $\ket{\psi}$ where $U_1, U_2 \in \{I, X, Y, Z\}$. In traditional superconducting architectures (our baseline), we consider a few error sources--storage error, one and two-qubit gate error, and measurement error. In superconducting architectures with resonant cavities such as our design, there is more nuance. We consider cavity storage and transmon storage error rates separately since each has its own coherence time and we separate transmon-transmon two-qubit gates and transmons-cavity two-qubit gates. We detail this and our other assumptions for simulation in experimental setup.

%
%
%

\begin{figure*}
    \centering
    \makebox[\textwidth][c]{
        \scalebox{\figscale}{%
        \scalebox{\figcnotextrascale}{%
        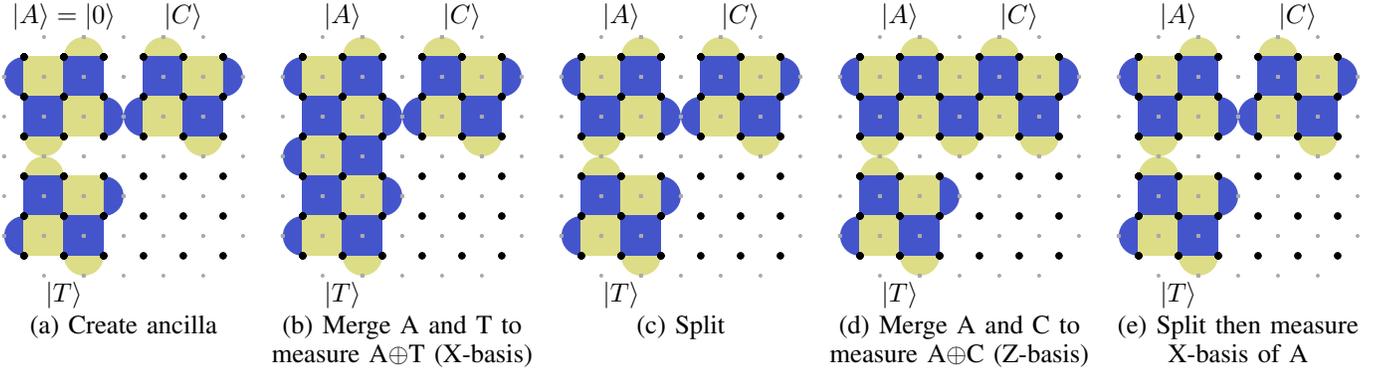\hspace{1.5em}}}
    }
    \caption{The lattice surgery operations to perform a logical CNOT on the standard surface code (and directly supported in our architecture).  Given control and target qubits $\ket{C}$ and $\ket{T}$, a CNOT is performed by enabling and disabling the parity checks as shown across 6 timesteps ((e) is two steps).  We show this complex process to contrast with the fast transversal CNOT enabled by our architecture (described later in Section \ref{sec:cnot}).
    }
    \label{fig:lattice-cnot}
\end{figure*}

\subsection{Surface Codes, Error Decoding, and \\Lattice Surgery}
\label{sec:qec}
The surface code \cite{fowler-braid} is one of the most promising quantum error correction protocols because it requires only nearest neighbor connectivity between physical qubits. The surface code is implemented on a two-dimensional array of physical qubits. These qubits are either data, where the state of the logical qubit is stored, or ancilla used for syndrome extraction (parity checks). These ancilla qubits are measured to stabilize the entangled state of the data. These ancilla fall into two categories, measure-Z and measure-X for Z syndromes and X syndromes designed to detect bit and phase errors respectively. Data qubits not on the boundary are adjacent to two measure-Z and two measure-X qubits.

In Figure~\ref{fig:regular-surface-code} we show four logical qubits with code distance 3 mapped to a 2D lattice of superconducting qubits. Dark physical qubits are used as data and light qubits are used as measure qubits. In this paper, we opt to explicitly indicate qubits in order to make clear how logical qubits, formed of many square and half-circle plaquettes, are mapped directly to hardware.  In our diagrams however, we use customary notation by shading X-plaquettes blue (dark) and Z-plaquettes yellow (light). Half-plaquettes contain only 2 data qubits and are shown as half circles.

Each X (Z) plaquette corresponds to a single measure-X (Z) qubit and the four data which it interacts with. The corners of each plaquette are the data qubits. For the baseline, we use standard Z and X syndrome extraction (parity measurement) circuits where the qubits of this circuit are physical qubits. The Z-syndrome measures the bit-parity of its corner qubits and the X-syndrome measures their phase-parity.  By repeatedly performing syndrome extraction and detecting parity changes we are able to locate errors. This repeated syndrome extraction collapses any error to a correctable Pauli error and forces the data to remain in what is called the code, or quiescent, state. Once the qubits are in this state, subsequent syndrome extraction should result in the same outcomes. If errors occur, we detect them as changes in measurement outcomes.

Errors are decoded by running a classical algorithm on the measured syndromes \cite{fowler-qec-matching}. In the surface code, when an error occurs on a data qubit, for example a single X bit-flip error, we see this as a change in the measurement outcome of \textit{both} of the Z-syndrome ancilla adjacent to it. If an error occurs on every data qubit in a chain of neighbors, only the two syndromes at the ends will detect a change.  The standard way of performing error decoding is to collect all of these changed syndromes into a complete graph with edge weights given by the log-probability of that chain of errors occurring. We perform a maximum likelihood perfect matching of this graph to find the most probable set of error locations which we correct or track in the classical control. If errors are sufficiently low these error chains will be well isolated and this decoding algorithm will be able to determine the correct set of corrections to be made. If errors are less sparse, this matching algorithm may misidentify which error chains have actually occurred and this can result in a logical error, that is a \textit{logical} bit flip or phase flip is applied. These logical errors cannot be detected because they result from misidentifying the physical errors.

There are two primary ways to manipulate the logical qubits of the surface code to perform desired logical operations--braiding and lattice surgery. In this paper we will primarily consider lattice surgery which has been shown to have some advantages over braiding like using fewer physical qubits. For a more thorough introduction to lattice surgery we refer the reader to \cite{fowler-lattice,game-of-codes,lao2018mapping}. In our proposed scheme, all primitive lattice surgery operations can be used such as split and merge which together perform a logical CNOT as shown in Figure \ref{fig:lattice-cnot}. For universal quantum computation in surface codes we allow for the creation and use of magic states such as $\ket{T}$ or $\ket{CCZ}$. These states are necessary because the T and CCZ operations cannot be done transversely (using physical gates on the data in parallel to reliably perform the logical gate) in this type of code. However, high fidelity versions of these states can be generated via distillation \cite{bravyi-distillation,game-of-codes} where many error-prone copies of the state are combined to generate the state with low error probability. Our scheme permits the use of these methods in the same way as other surface code schemes and also allows more efficient implementations.

\section{Virtualized Logical Qubits}
\label{sec:vlq}

In this section we describe in detail our proposed architecture, an embedding of the surface code which virtualizes logical qubits, saving over 10x in required number of transmons.  This takes advantage of quantum resonant cavity memory technology described above to store \textit{logical} qubits, in the form of surface code patches, in memory local to the computational transmons. In this section we describe how we can embed surface code tiles in two variations, Natural and Compact.  We show the hardware operations needed to perform efficient syndrome extraction for both in our new fault-tolerant architecture. We then describe how typical lattice surgery operations are translated into operations in this new scheme, and finally how our system supports fault-tolerant transversal interactions between logical qubits sharing the same virtual address. We verify these operations via process tomography. We briefly describe how magic state distillation, an important primitive for algorithms, is translated to our system.

\begin{figure}
    \centering
    \makebox[\linewidth][c]{%
        \scalebox{\figscale}{%
            $\vcenter{\hbox{\scalebox{1.1418309227}{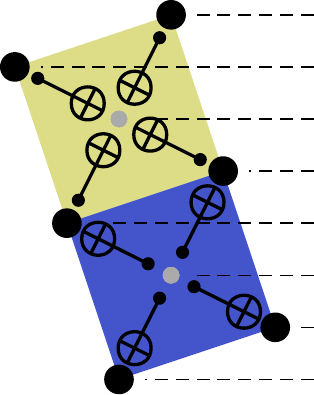}}}$%
            \hspace*{2.7pt}%
            $\vcenter{\hbox{\includegraphics[clip=false]{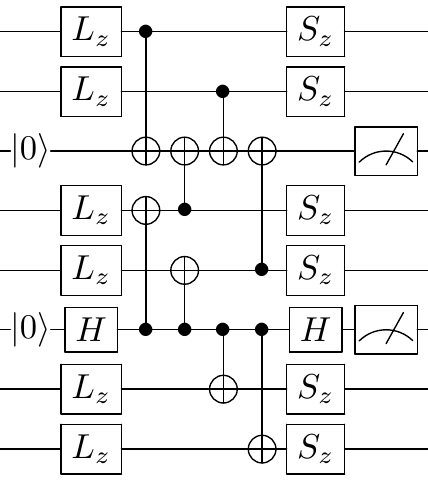}}}$%
        }%
    }
    \caption{
        Circuit showing how to execute our Natural embedding on hardware.  Left: The layout of eight data (black) and two ancilla (gray) in hardware.  CNOT operations between qubits are drawn between.  Right: A circuit diagram of the operations applied over time where each horizontal line corresponds to a qubit and each box or symbol is an operation.  The steps are $L_z$: load \added{from memory mode $z$}, $\ket{0}$: reset ancilla, CNOTs: compute the Z or X parity, Meter: measure the result, $S_z$: store back to memory.
    }
    \label{fig:zx-plaq-cnots}
\end{figure}

\subsection{Natural Surface Code Embedding}
\label{sec:natural}

Our goal here is to take logical qubits stored in a plane and find an embedding of that plane in 3D where the third dimension (our transmon-local memory) is a limited size, $k$.  The intuitive answer is to simply fold the surface $k$ times.  While this works, it does not have the benefits of a more clever embedding.  We propose slicing the plane into many pieces, storing them flat in memory to enable them to stitch together on-demand.  This embedding enables the fast transversal CNOT and high connectivity we will describe later.

Consider the high-level three dimensional view of the quantum memory architecture presented in \cite{schuster-arch}. For every transmon in this architecture (the compute qubits in the top layer of Figure~\ref{fig:stacked-surfaces}) there is a cavity attached with a fixed number of resonant modes, $k$. Each cavity can store $k$ qubits, one per mode. Each transmon can load and store qubits from its attached cavity by performing a transmon mediated iSWAP.
We assume all transmons can be operated on in parallel as is the case in most superconducting hardware (i.e. from IBM or Google). For example, we can load qubit $q_{iz}$ to transmon $t_i$ and load $q_{jz}$ to transmon $t_j$ in parallel, simultaneously execute single qubit operations on each qubit, then store in parallel. Any other qubits stored in cavities $i$ or $j$ will be unaffected by these operations.
We expect this technology to allow cavity size $k$ on the order of 10 to 100 qubits and it will likely not be practical to scale $k$ along with the size of the 2D grid as hardware improves so we cannot implement a true 3D code such as \cite{3d-code}. For our analysis, we conservatively assume $k=10$ and view this as a 2.5D architecture where we expect the width and height of the grid to scale while the depth, $k$, remains small.

We demonstrate how our system is sensitive to the length of these cavities in section \ref{sec:sensitivity-results} where the amount of time between error correction cycles is directly a function of this cavity size $k$. As the size of the cavity becomes very large, the physical qubits stored are expected to be subject to more and more decoherence errors which will reduce our ability to properly decode the errors.

Consider the rotated surface code of Figure~\ref{fig:regular-surface-code} and the high level view of this architecture in Figure~\ref{fig:stacked-surfaces}. We imagine mapping each of the physical qubits of this logical qubit $q_{L, 1}$ to the same mode $z$ of each cavity in this memory architecture. Another logical qubit $q_{L, 2}$ can be mapped to mode $z_2 \neq z$ of the same set of cavities. We view this as stacking the surface code patches, the logical qubits, together under the same set of transmon qubits. The transmons themselves are only used for logical operations and error correction cycles performed on the patches.

For logical qubits with code distance $d$ we define patches on the architecture, contiguous grids of size $d \times d$ data qubits and $d \times d$ ancilla qubits. Logical qubits are mapped to multiples of $d$ coordinates on the grid and a specific mode, $z$, for storage. For example, logical qubit $q_L$ is mapped to a pair $(P_{xy}, z)$ where $P_{xy}$ refers to the square patch of data transmons $q_{d\cdot x,d\cdot y}$ to $q_{d\cdot x+d-1,d\cdot y+d-1}$ and $z$ indicates which cavity mode it is stored in. A \textit{virtual memory address} of a logical qubit refers to exactly the pair (transmon patch, index). We sometimes refer to all pairs with the same transmon patch collectively as a stack where \textit{transmon patch} is the physical memory address where a patch is loaded.

In this memory architecture, recall we are unable to operate on qubits stored in the same cavity in parallel, however we \textit{are} permitted to operated on qubits stored in different cavities in parallel. This implies for two logical qubits $q_{L, 1}$ and $q_{L, 2}$ stored in the same stack we are only able to perform syndrome extraction on at most one of these qubits at a time. In order to detect measurement errors, we typically require $d$ rounds of syndrome extraction before we perform our decoding algorithm and correct errors. If all indices are occupied by logical qubits and we want to perform $d$ rounds of correction to each one we have two primary strategies. We can load a logical qubit (meaning load all data in parallel to each transmon), perform all $d$ rounds of extraction, then store the qubit.

Alternatively, we can Interleave the extraction cycles by loading the logical qubit in index 0, performing one syndrome extraction step, then storing. We execute this same procedure for every logical qubit in the stack and repeat $d$ times. We expect this latter procedure to be less efficient, subjecting the data qubits to $d$ load and store errors per $d$ cycles as opposed to performing exactly one set of loads and stores when collecting all $d$ measurements at once. We study the effect of this choice of syndrome extraction on the error threshold in Section \ref{sec:threshold-results}. We detail these extraction protocols for each syndrome in Figure~\ref{fig:zx-plaq-cnots}. Here we use $L_z$ ($S_z$) to indicate loading (storing) the data from (to) index $z$ of the attached cavity.

Intuitively, this scheme is stacking many different logical tiles together in a single location. This includes mapping measure-Z/X ancilla to cavity modes. However, this is unnecessary, because measure ancilla do not actually store any data and are reset before every extraction step. Therefore, we can reduce the number of cavities required for this system by simply omitting any cavity where ancilla are stored. Instead, every patch in the same stack shares the same ancilla, the transmons at the top layer with no attached cavity.

In our system, up to $k$ logical qubits share the same set of transmons, more efficiently storing these qubits than on a single large surface. In order to interact logical qubits in different stacks we load them in parallel to the transmons then interact them via lattice surgery operations like the CNOT shown in Figure~\ref{fig:lattice-cnot}. In these cases, all of the other stacks' transmons between the interacting logical qubits act as a single (possibly large) logical ancilla. In typical planar architectures, we are unable to execute transversal two-qubit operations due to limited connectivity. We can perform physical operations between qubits in the same cavity, mediated by the transmon. Therefore, in our system, we \textit{are} able to perform transversal two-qubit interactions if the logical qubits are co-located in the same stack.  We describe this next.

\subsection{Transversal CNOT}
\label{sec:cnot}

\begin{figure}[b]
    \centering
    \makebox[\linewidth][c]{
        \scalebox{\figscale}{
\begingroup%
  \makeatletter%
  \providecommand\color[2][]{%
    \errmessage{(Inkscape) Color is used for the text in Inkscape, but the package 'color.sty' is not loaded}%
    \renewcommand\color[2][]{}%
  }%
  \providecommand\transparent[1]{%
    \errmessage{(Inkscape) Transparency is used (non-zero) for the text in Inkscape, but the package 'transparent.sty' is not loaded}%
    \renewcommand\transparent[1]{}%
  }%
  \providecommand\rotatebox[2]{#2}%
  \newcommand*\fsize{\dimexpr\f@size pt\relax}%
  \newcommand*\lineheight[1]{\fontsize{\fsize}{#1\fsize}\selectfont}%
  \ifx\svgwidth\undefined%
    \setlength{\unitlength}{255.26073694bp}%
    \ifx\svgscale\undefined%
      \relax%
    \else%
      \setlength{\unitlength}{\unitlength * \real{\svgscale}}%
    \fi%
  \else%
    \setlength{\unitlength}{\svgwidth}%
  \fi%
  \global\let\svgwidth\undefined%
  \global\let\svgscale\undefined%
  \makeatother%
  \begin{picture}(1,0.65227424)%
    \lineheight{1}%
    \setlength\tabcolsep{0pt}%
    \put(0,0){\includegraphics[width=\unitlength,page=1]{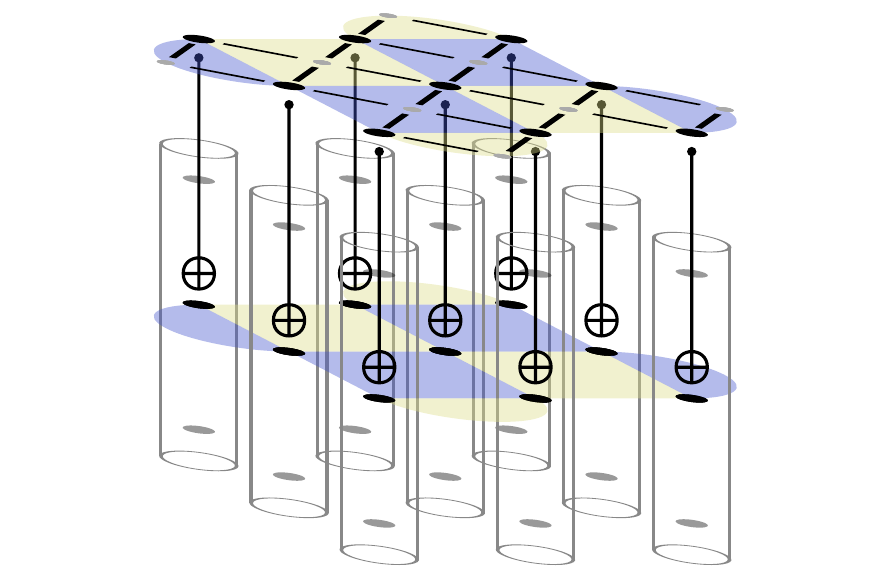}}%
    \put(0.79104318,0.41428228){\color[rgb]{0,0,0}\makebox(0,0)[lt]{\lineheight{1.25}\smash{\begin{tabular}[t]{l}CNOT gate\end{tabular}}}}%
    \put(0.17143546,0.29793067){\color[rgb]{0,0,0}\makebox(0,0)[rt]{\lineheight{1.25}\smash{\begin{tabular}[t]{r}mode $z$\end{tabular}}}}%
    \put(0,0){\includegraphics[width=\unitlength,page=2]{transversal-cnot.pdf}}%
    \put(0.87742544,0.56236617){\color[rgb]{0,0,0}\makebox(0,0)[lt]{\lineheight{1.25}\smash{\begin{tabular}[t]{l}logical\end{tabular}}}}%
    \put(0.87742544,0.5271081){\color[rgb]{0,0,0}\makebox(0,0)[lt]{\lineheight{1.25}\smash{\begin{tabular}[t]{l}control\end{tabular}}}}%
    \put(0,0){\includegraphics[width=\unitlength,page=3]{transversal-cnot.pdf}}%
    \put(0.87742544,0.2626726){\color[rgb]{0,0,0}\makebox(0,0)[lt]{\lineheight{1.25}\smash{\begin{tabular}[t]{l}logical\end{tabular}}}}%
    \put(0.87742544,0.22741452){\color[rgb]{0,0,0}\makebox(0,0)[lt]{\lineheight{1.25}\smash{\begin{tabular}[t]{l}target\end{tabular}}}}%
  \end{picture}%
\endgroup%
}
    }
    \caption{The transversal CNOT enabled by our 2.5D architecture.  The data qubits for the control logical qubit are loaded into the transmons.  Transmon-mediated CNOTs to mode $z$ for every data qubit perform the logical operation.  This takes one timestep to perform, 6x better than a lattice surgery CNOT.
    }
    \label{fig:transversal-cnot}
\end{figure}

\begin{figure*}[b]
    \centering
    \scalebox{\figscale}{%
    \scalebox{\figcompactextrascale}{%
\begingroup%
  \makeatletter%
  \providecommand\color[2][]{%
    \errmessage{(Inkscape) Color is used for the text in Inkscape, but the package 'color.sty' is not loaded}%
    \renewcommand\color[2][]{}%
  }%
  \providecommand\transparent[1]{%
    \errmessage{(Inkscape) Transparency is used (non-zero) for the text in Inkscape, but the package 'transparent.sty' is not loaded}%
    \renewcommand\transparent[1]{}%
  }%
  \providecommand\rotatebox[2]{#2}%
  \newcommand*\fsize{\dimexpr\f@size pt\relax}%
  \newcommand*\lineheight[1]{\fontsize{\fsize}{#1\fsize}\selectfont}%
  \ifx\svgwidth\undefined%
    \setlength{\unitlength}{445.50000161bp}%
    \ifx\svgscale\undefined%
      \relax%
    \else%
      \setlength{\unitlength}{\unitlength * \real{\svgscale}}%
    \fi%
  \else%
    \setlength{\unitlength}{\svgwidth}%
  \fi%
  \global\let\svgwidth\undefined%
  \global\let\svgscale\undefined%
  \makeatother%
  \begin{picture}(1,0.31360233)%
    \lineheight{1}%
    \setlength\tabcolsep{0pt}%
    \put(0,0){\includegraphics[width=\unitlength,page=1]{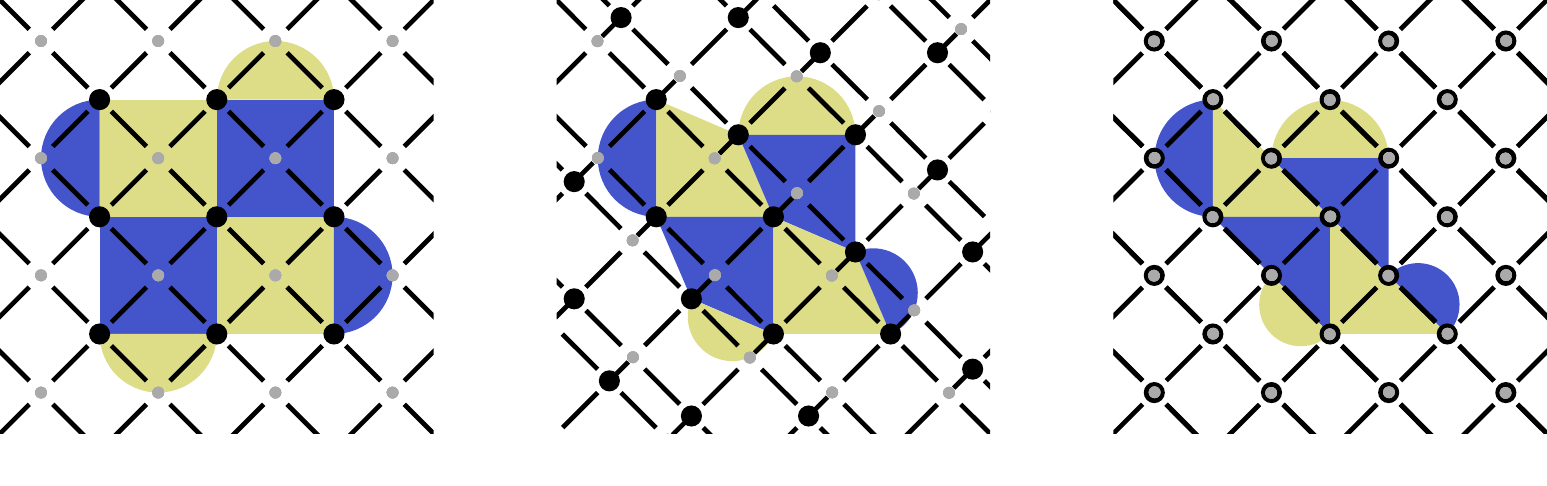}}%
    \put(0.14015152,0.00299627){\makebox(0,0)[t]{\lineheight{1.25}\smash{\begin{tabular}[t]{c}(a)\end{tabular}}}}%
    \put(0.5,0.00299627){\makebox(0,0)[t]{\lineheight{1.25}\smash{\begin{tabular}[t]{c}(b)\end{tabular}}}}%
    \put(0.85984848,0.00299627){\makebox(0,0)[t]{\lineheight{1.25}\smash{\begin{tabular}[t]{c}(c)\end{tabular}}}}%
    \put(0,0){\includegraphics[width=\unitlength,page=2]{compact-transform.pdf}}%
  \end{picture}%
\endgroup%
}}
    \caption{Transformation from Natural to Compact. (a)
    Natural embedding: Only data have attached cavities (not shown).
    (b) The transformation:
    Z ancilla (over yellow/light areas) merge with the upper-right data transmon and X ancilla (over blue/dark areas) merge with the lower-left data transmon.  The opposite parings are key to keeping 4-way grid connectivity.
    (c) Compact embedding: All ancilla transmons without attached cavities have been removed.  All remaining transmons have cavities and are used as both data and ancilla.
    }
    \label{fig:compact-transform}
\end{figure*}

A major advantage of this 2.5D architecture, enabled by our embedding of patches across memories, is the ability to do two-qubit operations transversely using the third dimension.  The logical operation is performed directly by doing the same physical gate to every data qubit and correcting any resulting errors. On typical 2D architecture error correcting codes like the surface code, the only transversal operations are single-qubit like X, Z, or H. Two-qubits operations are not possible because the corresponding data qubits of two logical patches cannot be made adjacent. However, with memory, it is possible to load one patch into the transmons and apply two-qubit gates mediated by each transmon onto the data qubits for a second qubit stored in one mode of the cavities. This works in both Natural and Compact (described later).

\begin{figure*}
    \centering
    \makebox[\textwidth][c]{
        \scalebox{\figscale}{%
        \scalebox{\figcnotextrascale}{%
        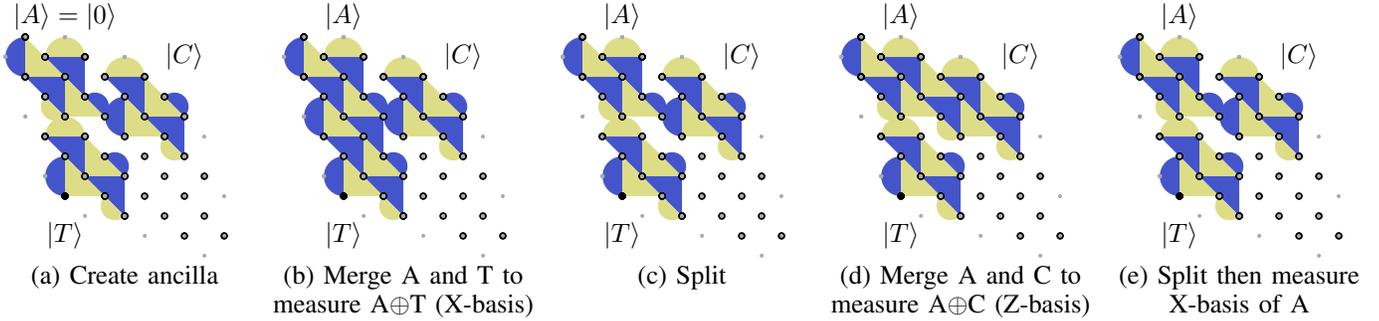\hspace{1.5em}}}
    }
    \caption{The Compact lattice surgery operations to perform a CNOT.  The logical operations performed are identical to Figure~\ref{fig:lattice-cnot} but the corresponding physical operations are arranged as shown in Figure~\ref{fig:compact-transform}.  This uses half as many transmons as Natural.  As before, it takes 6 timesteps of $d$ error correction cycles each.}
    \label{fig:lattice-cnot-compact}
\end{figure*}

Figure~\ref{fig:transversal-cnot} demonstrates this for the transversal CNOT gate which we verified via process tomography \cite{neeley2010,nielsen-chuang} to apply the expected CNOT unitary in simulation. This can be performed in a single round of $d$ error correction cycles while the lattice surgery CNOT shown in Figures \ref{fig:lattice-cnot} (and later \ref{fig:lattice-cnot-compact}) takes 6 rounds. This can translate to major savings in runtime for algorithms.

The transversal CNOT is not limited to logical qubits currently stored in the same 2D address.  With an extra step it is possible to transversely interact any two logical qubits. To do this one of the qubits must be \textit{moved} to the same 2D address as the other using a move operation described in \cite{game-of-codes}. The move operation involves growing the patch toward the move target in one step by adding new plaquettes along the entire path and performing $d$ cycles, one timestep, of error correction. Once grown, the patch can be shrunk from the other end back to its original size. The data qubits freed during the shrink are measured and used to determine any fixup operation. Once the two qubits are in the same 2D address, the transversal CNOT can be applied.  It can then be moved back, left where it is, or moved somewhere else as determined during compilation. This process takes 2 timesteps or 3 if including the second move.

\begin{figure}
    \centering
    \scalebox{\figscale}{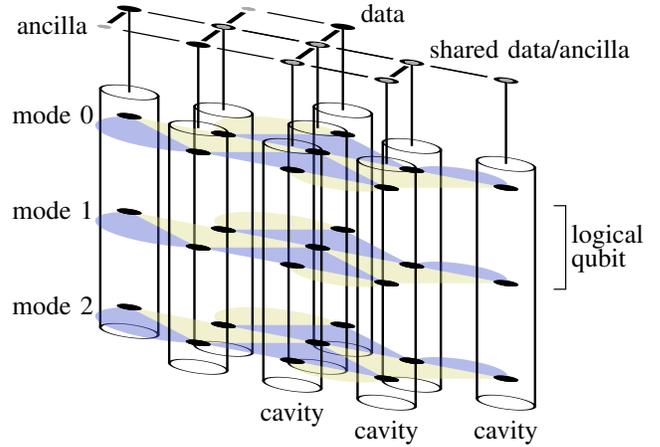}
    \caption{A 3D view of our Compact embedding.  Shown at the top is the 2D grid of transmon qubits.  Attached below every transmon is a resonant cavity.  Compact surface code patches are shown stored, one in each mode.  This deformed patch can be tiled in 2D.
    }
    \label{fig:stacked-compact-surfaces}
\end{figure}

\begin{figure*}
    \centering
        \scalebox{\figscale}{%
        \scalebox{0.9}{
            $\vcenter{\hbox{\scalebox{1.1418309227}{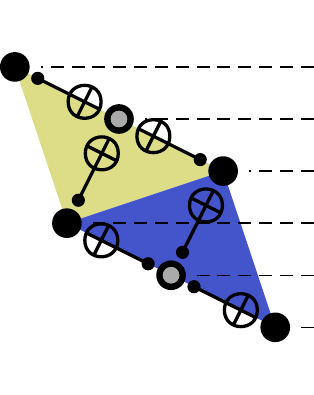}}}$%
            \hspace*{2.7pt}%
            $\vcenter{\hbox{\includegraphics[clip=false]{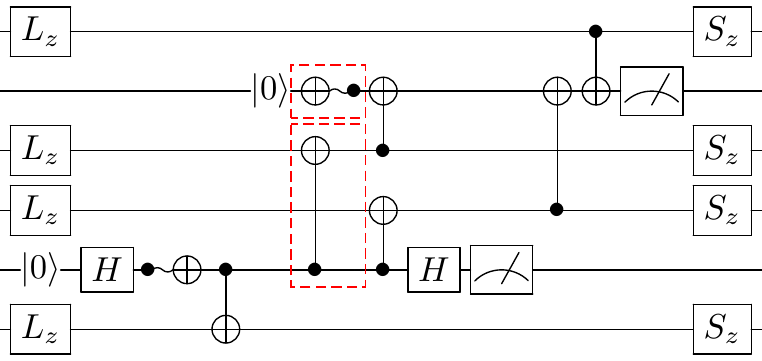}}}$%
            \hspace*{0.6cm}%
            $\vcenter{\hbox{\scalebox{0.77778}{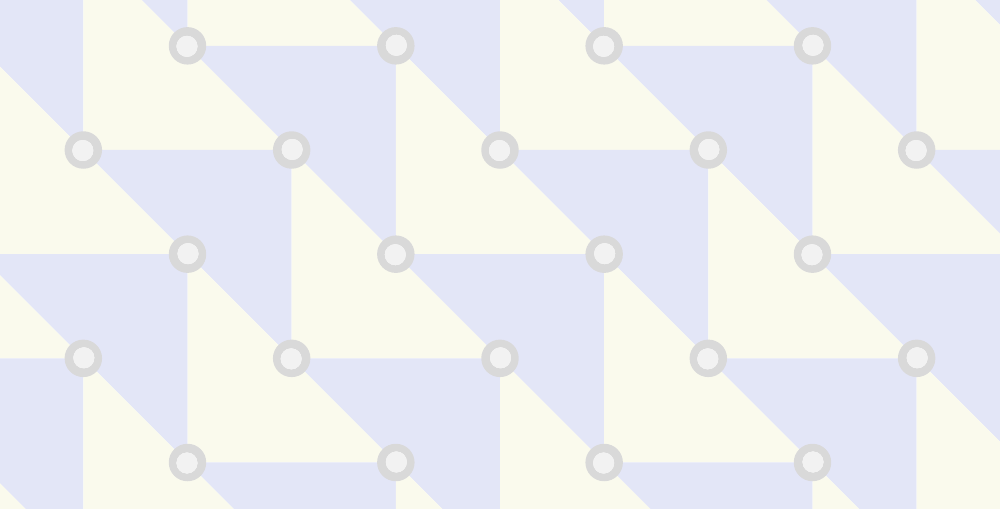}}}$%
        }}%
    \caption{The CNOT sequence for parity checks in Compact.  Left: A quantum circuit showing the hardware operations over time.  Right: The CNOT execution order repeats $A_0D_2$, $A_1D_3$, $A_2C_0$, $A_3C_1$, $B_0C_2$, $B_1C_3$, $B_2D_0$, $B_3D_1$.  The $AB$ and $CD$ sequences run in parallel but offset to ensure ancilla and data use do not conflict.  CNOTs for $A_0D_2$ are marked in red where an isolated circle indicates a transmon-mediated CNOT.
    }
    \label{fig:compact-plaq-order}
\end{figure*}

\subsection{Compact Surface Code Embedding}
\label{sec:compact}

In the previous scheme, half of the transmons did not have attached cavities (or they did not make use them). An ancilla and data qubit could share a transmon because the data are stored in the cavity the majority of the time and the ancilla are reset every cycle. This leads to a more efficient, Compact embedding which halves the required number of transmons. We will see that this comes at the cost of additional loads and stores from memory due to contention during error correction, effectively trading some error and time for significant space savings.

In the above memory architecture, because we do not store any logical qubits in the transmon layer, these qubits can act as the measurement ancilla, rather than have separate transmons only there to act as the syndrome measurement ancilla. With this observation, we can pack the data qubits of the surface code patch of Figure~\ref{fig:compact-transform}a more efficiently with \textit{every} transmon having a cavity attached. Each plaquette of the rotated surface code has a single ancilla at its center, interacting with each data qubit. For $Z$ plaquette (yellow or light) in this mapping scheme we colocate the upper-right data and the ancilla; the upper-right data is located in the cavity attached to the transmon corresponding to the ancilla. Similarly, for each $X$ plaquette (blue or dark) we colocate the lower-left data and the ancilla; the lower-left data is located in the cavity attached to the transmon corresponding to the ancilla.

This mapping results in plaquettes which resemble triangles rather than squares, where the center of the hypotenuse of each triangle corresponds to both the ancilla qubit and the data qubit, stored ``beneath'' in its cavity. Every data qubit is still mapped to the \textit{same} index. Notice in this scheme every data (sans the boundary) is still adjacent to two measure-Z and two measure-X ancilla where adjacent means either in the cavity of the ancilla or in a cavity adjacent to the ancilla. We illustrate this transformation from our undistorted Natural surface code patch to Compact in Figure~\ref{fig:compact-transform} and a diagram of this architecture with a cavity for every transmon in Figure~\ref{fig:stacked-compact-surfaces}.  If a different ancilla location were chosen, for example all sharing with the upper-right data, some of the syndrome extraction gates in the resulting arrangement would require six-way connectivity, two diagonal to the grid, which would be much more difficult to engineer with low noise. This scheme where X and Z ancilla share with data in opposite directions is the best scheme we found to satisfy the hardware connectivity.

In Natural, we assign square patches to predetermined square patches on the hardware. In Compact, we assign square patches to predetermined rhombus or diamond patches on the hardware. Previously, operations on the virtualized patches closely resembled the original operations because the shape was unchanged, except with the addition of loads and stores to retrieve the logical qubit from memory. The same operations apply here. We can examine the original, unmapped surface code patch and perform the same sequence of operations modulo loads and stores, on the transformed coordinates of the mapped version.

This new mapping also requires a new syndrome extraction procedure because data cannot be loaded while a transmon is in use as an ancilla. A single round of syndrome extraction can be executed by dividing the plaquettes into four groups, with each group containing non-interfering plaquettes.  Two plaquettes are non-interfering if they do not share their ancilla with any data qubits of the other plaquette. This process is detailed explicitly in Figure~\ref{fig:compact-plaq-order}. It is imperative this process use both the minimum number of loads and stores and keep data qubits loaded for as short a time as possible as the error incurred during this circuit directly impacts the error threshold for the code. This has a similar cost as Natural, Interleaved where a higher numbers of load and store gates were also required.

Error correction can be performed Interleaved or All-at-once just as with Natural.  This should be chosen dependent on how likely storage errors and gate errors are. For example, if storage errors are expected to be significant, we may opt to use Interleaved syndrome extraction. This will cost more loads and stores so if gate errors are more significant than storage errors we may opt for All-at-once.

\subsection{Architectural Considerations}
When compiling and executing programs in our system there are several important architectural features to keep in mind. First, it is always possible to execute a transversal two-qubit interaction, rather than requiring use of split and merge. In surface code architectures, the logical qubits are not bound to a specific hardware location and are free to move around on the grid. This qubit movement is fairly cheap requiring only a single round of $d$ error correction cycles (usually referred to as a single timestep) to move any distance. However, we require a clear area of unused patches to move through; typically, this requires about $1/3$ to $1/2$ of the total area to be kept as open channels to allow for distant qubit interactions. In our architecture this translates to keeping one of the resonant modes in every stack unused ($1/k$ of total qubits for cavity depth $k\approx 10$) and loading this mode along a path when a logical qubit needs to move, i.e there is an index in the stack which has no logical qubit mapped to it.  This enables our system to transport logical qubits between stacks to execute more time and space efficient transversal CNOTs. The empty mode is necessary for Compact because data is always stored back to the cavity during syndrome extraction but not required for Natural, All-at-once where the transmons themselves can act as the unused qubits to move the logical qubit through.

Unfortunately, this qubit movement is not entirely free. During the compilation process if we request many logical qubits to move in parallel this can be expensive due to serialization of intersecting move paths. Just as in current quantum systems without error correction where it is imperative to map and schedule multi-qubit interactions in a way which minimizes total execution time, it is also important in our system that logical qubits which interact heavily be located close by for similar reasons. The mapping problem on the system presented here is interesting because there is now a tradeoff between locality and serialization between operations with qubits sharing the same 2D address.

Second, we stress even though the logical qubits are stored in memory, they are still subject to errors and it is critical that every logical qubit be error corrected regularly. In the case of Interleaved syndrome extraction, every logical qubit of a stack will be roughly guaranteed to get a round of correction every $k$ time steps, where $k$ is the cavity depth. This rate is during steady state, when qubits are idle. When logical operations are being executed, this rate may be reduced slightly. When compiling and executing on this system, we may need to delay some operations in order to ensure stored logical qubits get the required amount of error correction and are not left so long that errors accumulate and error correction becomes less likely to succeed.

Finally, many lattice surgery operations require the use of ancilla logical qubits, for example to measure specific stabilizers which are done to execute a particular set of operations in \cite{game-of-codes}. This restriction requires our architecture and any compiler to guarantee one free mode of every stack be allocated to temporarily obtain large logical qubits. This free mode may be shared with qubit movement or separate if many ancilla logical qubits are used.

\section{Evaluation}
\label{sec:evaluation}

In this section, we outline our error model and experimental setup used to determine error thresholds for our mapping and syndrome extraction schemes. We compare to the surface code on a typical 2D architecture. Our goal is to demonstrate the error thresholds for various error correction schemes, i.e. to determine the necessary \textit{physical} error rate required to begin obtaining exponentially better \textit{logical} error rate as the code distance increases. Currently, neither transmon devices nor transmon-memory devices used for our schemes have consistently achieved physical error rates below this threshold and instead the threshold serves as a goal or checkpoint.

\subsection{Error Model and Noise Assumptions}
\label{sec:error-model}

For our experiments we make the following further assumptions about how noise and errors behave in both a typical 2D architecture and our 2.5D cavity memory architecture since both have the same fundamental underlying transmon technology:
\begin{itemize}
    \item The error rates in the device do not fluctuate appreciably over time.
    \item Transmon qubits can be actively reset and reinitialized to $\ket{0}$ efficiently and without significant error.
    \item All errors are independent.  No leakage errors and no correlated noise.
    \item All classical processing of the syndromes is instantaneous and error-free.
    \item Every $n$-qubit gate with the same $n$ is equally error-prone. For example, every one qubit operation has the exact same chance of failure regardless of which actual physical qubit it is applied to.
    \item All errors are Pauli, i.e. drawn from the set $\{I, X, Y, Z\}^{\otimes n}$. For example, if a one-qubit error occurs with probability $p$ then we apply an $X$, $Y$, or $Z$ with probability $p/3$ and $I$ (no error) with probability $1 - p$.
    \item We detect and correct $X$ and $Z$ errors independently.  A $Y$ error is both an $X$ and $Z$ error.
\end{itemize}

\begin{figure*}[p]
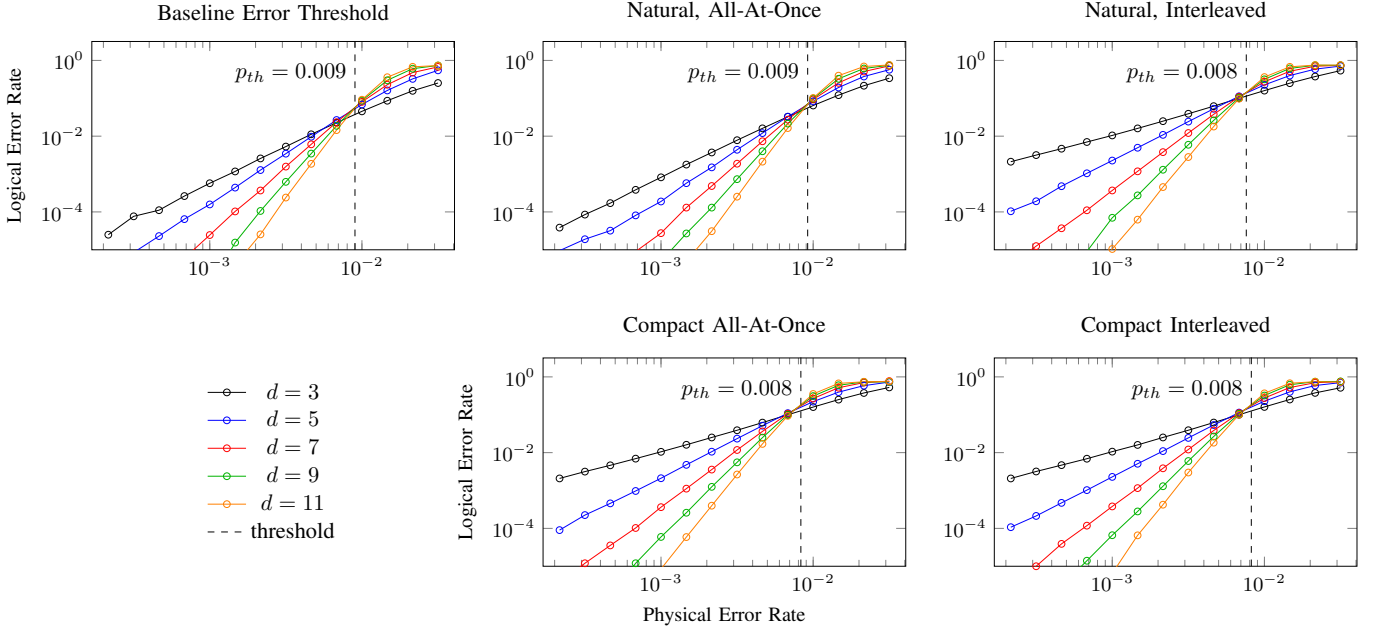

    \centering
    \scalebox{\figscaleplot}{%
    \makebox[0.99\textwidth][r]{%
        \input{figs/thresholds.tikz}%
    }}\\\vspace{1em}
    \scalebox{\figscaleplot}{%
    \makebox[0.99\textwidth][r]{%
        \input{figs/thresholds2.tikz}%
    }}\vspace{0.5em}
    \caption{Error thresholds for the baseline 2D architecture and Natural and Compact variants of our 2.5D architecture. The thresholds are comparable to the baseline indicating the space savings obtained in our system does not substantially reduce the error thresholds. The slopes of the lines in this figure indicate, post-threshold, how much improvement in physical error rates improve logical error rate.  Except for the baseline, all use a cavity size of 10.}
    \label{fig:base_thresholds}
\end{figure*}

For each of our experiments we rely on realistic device data for current superconducting devices, provided by IBM \cite{ibm-device}. For the memory hardware, we use experimental data from \cite{schuster-arch}. These parameters are listed in Table \ref{tab:gate_and_coherence_times}, where $T_{1, c}$ is the coherence time of the cavity, $T_{1, t}$ is the coherence time of the transmon, $\Delta_{t}$ is the single qubit gate time, $\Delta_{t-t}$ is the two-qubit transmon-transmon gate time, $\Delta_{t-m}$ is the two-qubit gate time of transmon-mode interactions, and $\Delta_{l/s}$ is the load and store times. In every experiment, the gate durations for one- and two-qubit interactions is fixed. In a first set of experiments, we vary all gate errors and coherence times together, all derived from a single probability of error $p$ given as the probability of an SC-SC (Transmon-Transmon gates) two-qubit gate error. We consider $T_1$ times of both cavities and transmons to determine the probability of storage error given as $\lambda = 1 - \exp{-\Delta t / T_1}$, where $\Delta t$ is the duration stored. We consider the same potential gate error rates for each of these devices since the underlying technology behaves very similarly. While typical coherence errors are not generally Pauli, we model them as Pauli errors here as a worst-case approximation since correcting Pauli errors is harder than correcting coherence errors in general.

\begin{table}
    \caption{Starting point coherence times and constant gate times \added{for the hardware models}.}
    \label{tab:gate_and_coherence_times}
    \centering
    \begin{tabular}{c|c|c}
        \begin{tabular}{c}Hardware\\Parameter\end{tabular}
        & \begin{tabular}{c}Baseline\\Transmons\end{tabular}
        & \begin{tabular}{c}Transmons\\with Memory\end{tabular}
        \\\hline\hline
        $T_{1,t}$ & 100 $\mu$s & 100 $\mu$s \\\hline
        $T_{1,c}$ & - & 1 ms \\\hline
        $\Delta_{t-t}$ & 200 ns & 200 ns \\\hline
        $\Delta_{t}$ & 50 ns & 50 ns \\\hline
        $\Delta_{t-m}$ & - & 200 ns \\\hline
        $\Delta_{l/s}$ & - & 150 ns \\
    \end{tabular}
\end{table}

\subsection{Experimental Setup}
\label{sec:experimental-setup}

In every experiment, we run 2,000,000 simulated trials per data point with each trial consisting of a round of error correction. We compute the logical error rate as the number of logical errors (misidentified error chains) over the total number of trials. The large number of trials is required to estimate logical error rates to $10^{-5}$. To determine the error threshold values for different surface code schemes, we vary the physical error rate over several different code sizes. The goal is to find an intersection point for each of these lines which gives a physical error rate below which we expect our logical error rate to get better as the physical error rate improves.  Below the threshold we also expect the logical error rate to get better exponentially in the code distance $d$.

We study 5 setups to determine initial error thresholds.
\begin{itemize}
    \item The surface code on a 2D superconducting architecture as our baseline.
    \item Our Natural embedding with either the All-at-once or Interleaved syndrome extraction.
    \item Our Compact embedding with either the All-at-once or Interleaved syndrome extraction.
\end{itemize}

In our designs, the possible sources of error are more nuanced and we study the thresholds' sensitivity to variation in the parameters.
In all threshold experiments, we assume cavity depth of 10 but later study sensitivity to cavity size.
\added{The simulation code used to generate our results is available on GitHub\cite{sourcecode}.}

\begin{figure*}[p]
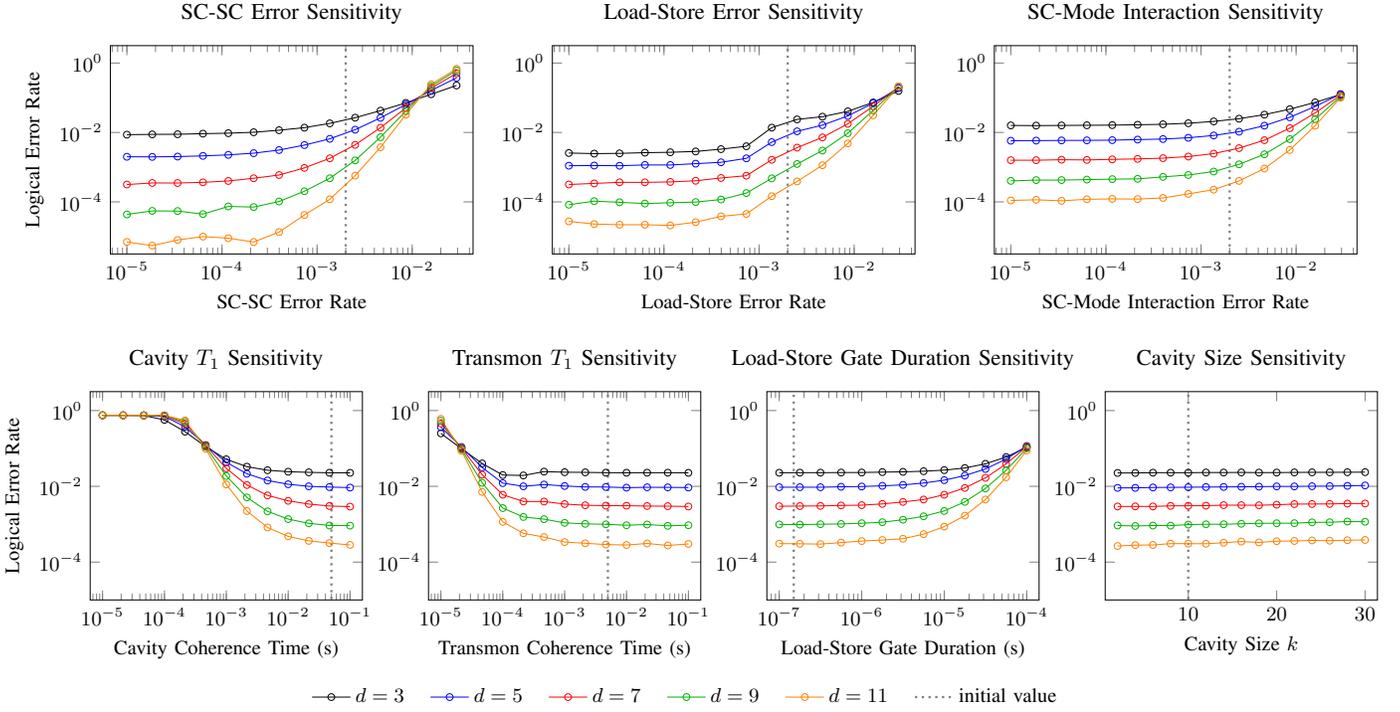

    \centering
    \scalebox{\figscaleplot}{%
    \makebox[0.99\textwidth][r]{%
        \input{figs/sensitivity_plots.tikz}%
    }}\\\vspace{1em}
    \scalebox{\figscaleplot}{%
    \makebox[1.01\textwidth][r]{%
        \input{figs/sensitivity_plots2.tikz}%
    }}\\\vspace{0.5em}
    \scalebox{\figscaleplot}{%
    \makebox[\textwidth][c]{%
        \begin{tikzpicture}[baseline,scale=0.85,trim axis left,trim axis right]
\pgfplotsset{every tick label/.append style={font=\small}}
\pgfplotsset{every axis label/.append style={font=\small}}
    \begin{axis}[
        name=plot2,
        title={  },
        xlabel={},
        ylabel={},
        width={0.8\textwidth},
        height={50pt},
        legend style={
            draw=none,
            at={(0.5,0)},
            anchor=south,
            font=\small},
        legend columns=-1,
        axis line style={draw=none},
        tick style={draw=none},
        xticklabels={},
        yticklabels={},
        xmin=-2, xmax=-1, ymin=-2, ymax=-1, clip=true,
    ]
        \addplot[color=black, mark=o, mark size=1.5pt] coordinates{(0, 0)};
        \addlegendentry{$d=3$~~~~}
        \addplot[color=blue, mark=o, mark size=1.5pt] coordinates{(0, 0)};
        \addlegendentry{$d=5$~~~~}
        \addplot[color=red, mark=o, mark size=1.5pt] coordinates{(0, 0)};
        \addlegendentry{$d=7$~~~~}
        \addplot[color=green!70!black, mark=o, mark size=1.5pt] coordinates{(0, 0)};
        \addlegendentry{$d=9$~~~~}
        \addplot[color=orange, mark=o, mark size=1.5pt] coordinates{(0, 0)};
        \addlegendentry{$d=11$~~~~}
        \addplot[color=gray, mark=none, dotted, line width=1pt] coordinates{(0, 0)};
        \addlegendentry{\added{initial value}}
    \end{axis}
\end{tikzpicture}%
    }}
    \caption{Sensitivity of logical error rate to various error sources in Compact, Interleaved. The logical error \added{rates are} most sensitive to physical error of Loads/Stores and SC-SC gates. The logical error rate is \added{less sensitive to the coherence times and} mostly insensitive to effects of \added{load-store duration and} cavity size.}
    \label{fig:sens_results}
\end{figure*}

\section{Error Threshold Results}
\label{sec:threshold-results}

We detail our threshold results in Figure~\ref{fig:base_thresholds}. We study 5 different code distances in order to obtain the physical error threshold value. The threshold value indicates at which point increasing the code distance, $d$, improves the logical error rate instead of hurting it. This threshold is a function of both the physical system model, the chosen syndrome extraction circuit, and the specific decoding procedure. For example, decoding procedures which do not accurately represent the probability of certain error chains occurring will do a poor job of correcting those errors. The decoding process should be directly informed by the error model. In systems with more complicated error models, the decoder should be aware of these further details to inform its decision about which types of errors occurred and the proper way to correct them. We use the usual maximum likelihood decoder because we use standard assumptions in our error model.

The major difference in each procedure is the additional error sources and different syndrome extraction procedures. For example, the baseline is not subject to any of the effects related to cavity storage or transmon-mode operations. These syndrome extraction procedures differ by the amount of storage time of data qubits in different locations (cavity vs. transmon) as well as the number of different physical gate operations applied to them. These differences however, do not cause substantial variation in the error threshold for the different setups which is extremely promising. Second, the slopes for each code distance compared across the various schemes is stable, indicating each scheme improves at a similar rate, post error threshold, and showing that the logical error rate decays exponentially with $d$ as desired. This is significant because it means we will be able to save on total number of transmons without major shifts in the error threshold. Since transmon memory technology is expected to perform as well as other competing transmon technology, we obtain higher distance codes, and hence better logical error rate, with fewer total transmons.

\section{Error Sensitivity Results}
\label{sec:sensitivity-results}

In this section, we study the effects of different sources of error on the thresholds obtained in section \ref{sec:threshold-results}. Specifically, we show how different system-level details affect the threshold of the code. Here we focus on Compact, Interleaved as the most efficient physical qubit mapping and subject to a wide variety of errors. In these studies, the physical error rates of all but a single error source are fixed at a typical operating point below the threshold obtained previously, $2\times 10^{-3}$ and the cavity depth is fixed at 10. \added{Gate} times are fixed \added{while} we vary the physical error rate of SC-SC gates, SC-Cavity gates, Load-Store gates \added{or} the coherence times of the cavity and the transmon. \added{We additionally study the duration of load/store, the gates unique to memory technology. We note the effect of the SC-Cavity gate duration will be a similar, smaller effect since it occurs only once per qubit per error correction cycle.} Finally, we study the effect of cavity size by varying the number of modes per cavity, \added{causing a proportional} delay between error correction cycles.

The results of these sensitivity studies are found in Figure~\ref{fig:sens_results}. The logical error rate is sensitive to a particular error source's probability if the slope of the line is pronounced \added{at the marked reference value}. The logical error rate for Compact, Interleaved \added{is sensitive to all changes in system-level details to some degree. The gate error rates show the highest sensitivity, indicating improvement in these will give the greatest benefit.
Coherence times are not quite as sensitive but the slightly over 10x offset between the cavity and transmon plots shows that there is no benefit in transmon $T_1$ being longer than $1/10$ cavity $T_1$ when the cavity size is 10.}
\added{The} lines taper off, indicating other errors sources \added{eventually dominate}. Initially, we expected the cavity size to have a large impact on the logical error rate. However, when coherence times are high and gate error rates are fairly low below the threshold, the logical error rate does increase proportional to the length of the cavity but the effect is very minor. This indicates, given cavities with good coherence times, our proposed system will be able to scale smoothly into the future as cavity sizes increase.

While larger cavity sizes will make this architecture even more advantageous, there will be a point at which it has a vanishing benefit because the delay between error correction becomes too long and decoherence error dominates.  For the error rates used in the evaluation, we find that cavity decoherence error starts dominating after cavity size $k\approx 150$.  After this point, it would be more beneficial to improve cavity coherence time.
\vspace{6pt}  

\begin{figure}[t]
    \centering
    \scalebox{\figscaleplot}{%
    \makebox[\linewidth][l]{%
        \begin{tikzpicture}
  \begin{axis}[
    ybar,
    ylabel={$\ket{T}$ Production Rate},
    title={(a) Rate with 100 Patches~~~~~},
    symbolic x coords={Fast, Small, VQubits},
    x tick label style={rotate=45,anchor=east},
    width=0.5*\linewidth,
    height=0.42*\linewidth,
    xtick style={draw=none},
    ytick pos=left,
    enlarge x limits=0.35,
    ymin=0,
    ]
    \addplot coordinates {(Fast, .555)(Small, .826)(VQubits, 1.010)};
  \end{axis}
\end{tikzpicture}%
\hspace{-2em}
\begin{tikzpicture}
  \begin{axis}[
    ybar,
    title={(b) Space To get 1 $\ket{T}$ / Step~~~~~},
    ylabel={\# Patches},
    symbolic x coords={Fast, Small, VQubits},
    x tick label style={rotate=45,anchor=east},
    width=0.5*\linewidth,
    height=0.42*\linewidth,
    xtick style={draw=none},
    ytick pos=left,
    enlarge x limits=0.35,
    ymin=0,
    ]
    \addplot coordinates {(Fast, 180)(Small, 121)(VQubits, 99)};
  \end{axis}
\end{tikzpicture}%
    }}\\
    \vspace{-1em}
    \caption{(a) The T-state generation rates of three different circuits.  Higher generation rate is better.  (b) The space, in terms of number of patches, required to produce a single $\ket{T}$ per time step. Lower is better.  Fast \cite{fast-distillation} and Small \cite{game-of-codes} work in the surface code and do not use memory. VQubits is implemented with transversal CNOTs in our 2.5D architecture. All are based on \cite{bravyi-distillation}.}
    \label{fig:distilation-rate}
\end{figure}
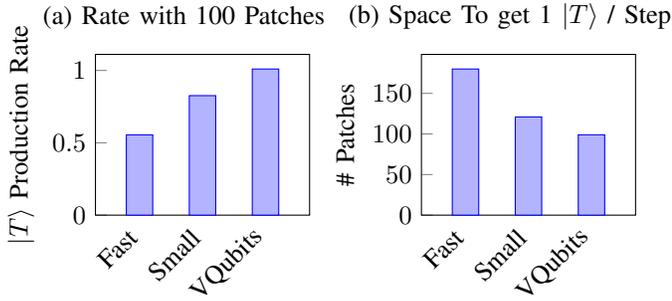

\section{Magic State Distillation \\Resource Estimates}
\label{sec:resource-estimates}

Now that we have shown error correction is effective in our virtualized qubit architecture, we analyze how the transversal CNOT and memory connectivity can benefit the performance of an algorithm overall.  In error-corrected quantum algorithms, the dominating cost (commonly $>$ 90\%) in both space and time resources is magic state distillation \cite{other-magic-estimates,adam-magic-estimates,gidney2019factor}.  For this analysis we consider how T-state distillation, a commonly used magic state, is improved.  Any improvements here will translate directly to improvements in important algorithms like Shor's and Grover's.

We take the 15-to-1 distillation circuit of \cite{bravyi-distillation} to generate a T magic state using a single patch of transmons with 6 logical qubits stored in the attached cavities.  This circuit consists of 16 qubit initializations, 15 measurements, 35 CNOT gates and a few other operations.  It takes a total of 110 surface code timesteps to generate a T-state using only a single patch of transmons.  If pairs of these circuits are executed in lock-step, they only take 99 timesteps.

In Figure~\ref{fig:distilation-rate} we compare the T-state generation rate with memory against two representative extremes designed for
\newpage 
speed or size, Fast Lattice \cite{fast-distillation} and Small Lattice \cite{game-of-codes} \added{(also based on \cite{bravyi-distillation})}.  Fast Lattice generates a T-state every 6 timesteps but uses 30 patches of space whereas Small Lattice, generates a T-state every 11 timesteps using only 11 patches of space.  We compare these results by computing the T-state generation rate per timestep if we filled 100 patches with copies of the circuit running in parallel.
Table \ref{tab:magic-cost} show the qubit cost of each and chip area will be proportional to the number of transmons.  Using our VQubits protocol generates 1.82x as many T-states as Fast Lattice and 1.22x as many as Small Lattice.
This improvement allows an algorithm like Shor's to run roughly 1.22x faster or work on smaller hardware.

\begin{table}[t]
    \caption{Transmon, depth-10 cavity, and total qubit costs of each T-state generation protocol for $d=5$.}
    \label{tab:magic-cost}
    \centering
    \small
    \begin{tabular}{c|c|c|c}
        Protocol & \# transmons 
        & \# cavities & total qubits
        \\\hline\hline
        Fast Lattice \cite{fast-distillation} & 1499 & - & 1499 \\
        Small Lattice \cite{game-of-codes} & 549 & - & 549 \\
        \textbf{VQubits (natural)} & \textbf{49} & \textbf{25} & \textbf{299} \\
        \textbf{VQubits (compact)} & \textbf{29} & \textbf{25} & \textbf{279} \\
    \end{tabular}
\end{table}
\section{Conclusion}
\label{sec:conclusion}

Realizable quantum error correction protocols are a critical step in the path towards fault-tolerant quantum computing. There has been great progress in NISQ-era devices, but it is equally critical to look towards designing architectures for QEC. In this paper, we introduce a system which virtualizes logical, error corrected qubits and is both space and time efficient without sacrificing in terms of fault tolerance. 

By taking advantage of recent advances in quantum memory technology, we present a new architecture to substantially reduce hardware requirements by storing logical qubits distributed in memory. This technology allows memory to be separated but local to computation in a quantum system. We provide two direct mappings of the surface code to this new system with virtual addressing and illustrate how syndrome extraction and error correction procedures can be executed efficiently on the embedded surface code. Our embedding, combined with the random-access nature of the memory is important for several reasons. It enables fast transversal gates like the CNOT which can reduce program execution time by allowing faster operations and indirectly through improved magic-state distillation protocols. It significantly reduces the total number of transmon qubits required (10x for our analysis) which allows larger code distance patches while using 10x fewer transmon qubits and classical control wires.  This allows error correction to be realized much sooner on small architectures.  Our results show superconducting cavity-based architectures offer a promising path towards quickly scaling fault-tolerant quantum computation and can be evaluated with 10 logical qubits using as few as 11 transmons and 9 cavities.  We hope this work motivates further experimental efforts and prompts industry to adopt and scale-up this architecture.

\newpage


\begin{thebibliography}{10}
\providecommand{\url}[1]{#1}
\csname url@samestyle\endcsname
\providecommand{\newblock}{\relax}
\providecommand{\bibinfo}[2]{#2}
\providecommand{\BIBentrySTDinterwordspacing}{\spaceskip=0pt\relax}
\providecommand{\BIBentryALTinterwordstretchfactor}{4}
\providecommand{\BIBentryALTinterwordspacing}{\spaceskip=\fontdimen2\font plus
\BIBentryALTinterwordstretchfactor\fontdimen3\font minus
  \fontdimen4\font\relax}
\providecommand{\BIBforeignlanguage}[2]{{%
\expandafter\ifx\csname l@#1\endcsname\relax
\typeout{** WARNING: IEEEtran.bst: No hyphenation pattern has been}%
\typeout{** loaded for the language `#1'. Using the pattern for}%
\typeout{** the default language instead.}%
\else
\language=\csname l@#1\endcsname
\fi
#2}}
\providecommand{\BIBdecl}{\relax}
\BIBdecl

\bibitem{ibm-device}
\BIBentryALTinterwordspacing
``Quantum devices \& simulators,'' Jun 2018. [Online]. Available:
  \url{https://www.research.ibm.com/ibm-q/technology/devices/}
\BIBentrySTDinterwordspacing

\bibitem{nisq}
J.~Preskill, ``Quantum computing in the nisq era and beyond,'' \emph{Quantum},
  vol.~2, p.~79, 2018.

\bibitem{quantum-supremacy}
\BIBentryALTinterwordspacing
F.~Arute, K.~Arya, R.~Babbush, D.~Bacon, J.~C. Bardin, R.~Barends, R.~Biswas,
  S.~Boixo, F.~G. S.~L. Brandao, D.~A. Buell, B.~Burkett, Y.~Chen, Z.~Chen,
  B.~Chiaro, R.~Collins, W.~Courtney, A.~Dunsworth, E.~Farhi, B.~Foxen,
  A.~Fowler, C.~Gidney, M.~Giustina, R.~Graff, K.~Guerin, S.~Habegger, M.~P.
  Harrigan, M.~J. Hartmann, A.~Ho, M.~Hoffmann, T.~Huang, T.~S. Humble, S.~V.
  Isakov, E.~Jeffrey, Z.~Jiang, D.~Kafri, K.~Kechedzhi, J.~Kelly, P.~V. Klimov,
  S.~Knysh, A.~Korotkov, F.~Kostritsa, D.~Landhuis, M.~Lindmark, E.~Lucero,
  D.~Lyakh, S.~Mandr{\`a}, J.~R. McClean, M.~McEwen, A.~Megrant, X.~Mi,
  K.~Michielsen, M.~Mohseni, J.~Mutus, O.~Naaman, M.~Neeley, C.~Neill, M.~Y.
  Niu, E.~Ostby, A.~Petukhov, J.~C. Platt, C.~Quintana, E.~G. Rieffel,
  P.~Roushan, N.~C. Rubin, D.~Sank, K.~J. Satzinger, V.~Smelyanskiy, K.~J.
  Sung, M.~D. Trevithick, A.~Vainsencher, B.~Villalonga, T.~White, Z.~J. Yao,
  P.~Yeh, A.~Zalcman, H.~Neven, and J.~M. Martinis, ``Quantum supremacy using a
  programmable superconducting processor,'' \emph{Nature}, vol. 574, no. 7779,
  pp. 505--510, 2019. [Online]. Available:
  \url{https://doi.org/10.1038/s41586-019-1666-5}
\BIBentrySTDinterwordspacing

\bibitem{shor}
\BIBentryALTinterwordspacing
P.~W. Shor, ``Polynomial-time algorithms for prime factorization and discrete
  logarithms on a quantum computer,'' \emph{SIAM Journal on Computing},
  vol.~26, no.~5, pp. 1484--1509, Oct. 1997. [Online]. Available:
  \url{http://dx.doi.org/10.1137/S0097539795293172}
\BIBentrySTDinterwordspacing

\bibitem{grover}
L.~K. Grover, ``A fast quantum mechanical algorithm for database search,'' in
  \emph{Annual ACM Symposium on Theory of Computing}.\hskip 1em plus 0.5em
  minus 0.4em\relax ACM, 1996, pp. 212--219.

\bibitem{color-code}
A.~J. Landahl, J.~T. Anderson, and P.~R. Rice, ``Fault-tolerant quantum
  computing with color codes,'' \emph{arXiv preprint arXiv:1108.5738}, 2011.

\bibitem{fowler-braid}
A.~G. Fowler, M.~Mariantoni, J.~M. Martinis, and A.~N. Cleland, ``Surface
  codes: Towards practical large-scale quantum computation,'' \emph{Physical
  Review A}, vol.~86, no.~3, p. 032324, 2012.

\bibitem{fowler-lattice}
C.~Horsman, A.~G. Fowler, S.~Devitt, and R.~Van~Meter, ``Surface code quantum
  computing by lattice surgery,'' \emph{New Journal of Physics}, vol.~14,
  no.~12, p. 123011, 2012.

\bibitem{gidney2019factor}
C.~Gidney and M.~Eker{\aa}, ``How to factor 2048 bit rsa integers in 8 hours
  using 20 million noisy qubits,'' \emph{arXiv preprint arXiv:1905.09749},
  2019.

\bibitem{martinis-caltech}
\BIBentryALTinterwordspacing
J.~Martinis, ``Quantum supremacy using a programmable superconducting
  processor,'' 11 2019, institute for Quantum Information and Matter Seminar at
  the California Institute of Technology. [Online]. Available:
  \url{https://youtu.be/FklMpRiTeTA}
\BIBentrySTDinterwordspacing

\newpage

\bibitem{schuster-arch}
R.~Naik, N.~Leung, S.~Chakram, P.~Groszkowski, Y.~Lu, N.~Earnest, D.~McKay,
  J.~Koch, and D.~Schuster, ``Random access quantum information processors
  using multimode circuit quantum electrodynamics,'' \emph{Nature
  communications}, vol.~8, no.~1, p. 1904, 2017.

\bibitem{game-of-codes}
D.~Litinski, ``A game of surface codes: Large-scale quantum computing with
  lattice surgery,'' \emph{Quantum}, vol.~3, p. 128, 2019.

\bibitem{nielsen-chuang}
M.~A. Nielsen and I.~L. Chuang, ``Quantum information and quantum
  computation,'' \emph{Cambridge: Cambridge University Press}, vol.~2, no.~8,
  p.~23, 2000.

\bibitem{yale-arch}
C.~T. Hann, C.-L. Zou, Y.~Zhang, Y.~Chu, R.~J. Schoelkopf, S.~M. Girvin, and
  L.~Jiang, ``Hardware-efficient quantum random access memory with hybrid
  quantum acoustic systems,'' \emph{arXiv preprint arXiv:1906.11340}, 2019.

\bibitem{fowler-qec-matching}
A.~G. Fowler, A.~C. Whiteside, and L.~C. Hollenberg, ``Towards practical
  classical processing for the surface code,'' \emph{Physical review letters},
  vol. 108, no.~18, p. 180501, 2012.

\bibitem{lao2018mapping}
L.~Lao, B.~van Wee, I.~Ashraf, J.~van Someren, N.~Khammassi, K.~Bertels, and
  C.~G. Almudever, ``Mapping of lattice surgery-based quantum circuits on
  surface code architectures,'' \emph{Quantum Science and Technology}, vol.~4,
  no.~1, p. 015005, 2018.

\bibitem{bravyi-distillation}
S.~Bravyi and J.~Haah, ``Magic-state distillation with low overhead,''
  \emph{Physical Review A}, vol.~86, no.~5, p. 052329, 2012.

\bibitem{3d-code}
H.~Bomb{\'\i}n, ``Gauge color codes: optimal transversal gates and gauge fixing
  in topological stabilizer codes,'' \emph{New Journal of Physics}, vol.~17,
  no.~8, p. 083002, 2015.

\bibitem{neeley2010}
M.~G. Neeley, ``Generation of three-qubit entanglement using josephson phase
  qubits,'' Ph.D. dissertation, University of California, Santa Barbara, 2010.

\bibitem{sourcecode}
C.~Duckering and J.~M. Baker, ``Simulation source code for virtualized logical
  qubits,'' \url{https://github.com/cduck/VLQ}, 2020.

\bibitem{fast-distillation}
D.~Litinski, ``Magic state distillation: Not as costly as you think,''
  \emph{arXiv preprint arXiv:1905.06903}, 2019.

\bibitem{other-magic-estimates}
S.~S. Tannu, Z.~A. Myers, P.~J. Nair, D.~M. Carmean, and M.~K. Qureshi,
  ``Taming the instruction bandwidth of quantum computers via hardware-managed
  error correction,'' in \emph{2017 50th Annual IEEE/ACM International
  Symposium on Microarchitecture (MICRO)}.\hskip 1em plus 0.5em minus
  0.4em\relax IEEE, 2017, pp. 679--691.

\bibitem{adam-magic-estimates}
Y.~Ding, A.~Holmes, A.~Javadi-Abhari, D.~Franklin, M.~Martonosi, and F.~Chong,
  ``Magic-state functional units: Mapping and scheduling multi-level
  distillation circuits for fault-tolerant quantum architectures,'' in
  \emph{2018 51st Annual IEEE/ACM International Symposium on Microarchitecture
  (MICRO)}.\hskip 1em plus 0.5em minus 0.4em\relax IEEE, 2018, pp. 828--840.

\end{thebibliography}


\end{document}